\documentclass[12pt]{article}
\usepackage{amssymb}
\usepackage{amsmath}
\usepackage{latexsym}
\usepackage[usenames]{color}
\usepackage{cite}
\usepackage{setspace}
\usepackage{graphicx} 
\usepackage{epsfig}

\usepackage[setpagesize=false,pagebackref=false, linktocpage, bookmarksopen=true, colorlinks=true, linkcolor=blue,citecolor=blue,urlcolor=blue]{hyperref}
\begin{document}
\vspace{5cm}
\begin{center}
{\LARGE\text{
Liouville Theory, AdS$_2$ String, and Three-Point Functions
}\par}

\vspace{2cm}

{\large \sc Shota Komatsu$^{\ast}$}
\\ \vspace{1.5cm}
\text{
$^{\ast}$School of Natural Sciences, Institute for Advanced Study, Princeton, New Jersey 08540, USA
}
\vspace{1cm}

{\tt shota.komadze AT gmail.com}

\par\vspace{3cm}

\textbf{Abstract}\vspace{2mm}
\end{center}
\noindent
This is a write-up of the lectures given in Young Researchers Integrability School 2017. The main goal is to explain the connection between the ODE/IM correspondence and the classical integrability of strings in AdS. As a warm up, we first discuss the classical three-point function of the Liouville theory. The starting point is the well-known fact that the classical solutions to the Liouville equation can be constructed by solving a Schr\"{o}dinger-like differential equation. We then convert it into a set of functional equations using a method similar to the ODE/IM correspondence. The classical three-point functions can be computed directly from these functional equations, and the result matches with the classical limit of the celebrated DOZZ formula. We then discuss the semi-classical three-point function of strings in AdS$_2$ and show that one can apply a similar idea by making use of the classical integrability of the string sigma model on AdS$_2$. The result is given in terms of the ``massive'' generalization of Gamma functions, which show up also in string theory on pp-wave backgrounds and the twistorial generalization of topological string.

\setcounter{page}{1}
\renewcommand{\thefootnote}{\arabic{footnote}}
\setcounter{footnote}{0}
\setcounter{tocdepth}{3}
\newpage

\tableofcontents

\parskip 5pt plus 1pt   \jot = 1.5ex
\section{Preface}
Recent years have witnessed significant development in the computation of various observables in planar $\mathcal{N}=4$ supersymmetric Yang-Mills theory \cite{review}. The driving force of this astonishing progress was the integrable structure discovered both in the gauge theory and in the dual string theory. On the gauge theory side, it emerged as the quantum integrability of the spin chain which describes the dynamics of the single-trace operator, whereas on the string theory side it appeared as the classical integrability of the string sigma model. This endeavor, initiated in the study of the spectrum of the operators, was recently extended to include more complicated observables such as the null polygonal Wilson loop and correlation functions, which contain richer dynamical information.

To compute these complicated observables on the string theory side, one has to study the classical string configuration with nontrivial topology, such as the surface ending on the null polygons or the surface with multiple legs. Although simple as a concept, it is hardly possible to find such complicated surfaces in practice. Nevertheless, progress can be made by the ingenious use of the integrability: Surprisingly, one can directly compute these observables without knowing the explicit shapes of the surfaces \cite{Y-system,JW,KKN,CT}\footnote{See also earlier related works on this subject \cite{KK4,KK3,KK2}.}.

The method employed in this procedure turned out to have close resemblance to a seemingly unrelated subject, called the ODE/IM correspondence \cite{ODE/IM}. The ODE/IM correspondence relates certain Schr\"{o}dinger equations to the functional equations which show up in the study of quantum integrable systems\footnote{See also the recent beautiful paper \cite{Marino}, which generalized the ODE/IM correspondence to a general class of potentials.}. Besides being of academic interest, it provides an efficient way of computing the spectrum of anharmonic oscillators and gives us insight into its non-perturbative structure. The techniques applied to classical strings can be regarded as the generalization of this correspondence, and the main goal of this lecture is to clarify their connections by taking the three-point function as an illustrative example.

With this aim in mind, we first discuss the three-point function in the classical Liouville theory in \hyperref[lec1]{Lecture I}. The goal of this part of the lecture is to explain the relation between the Liouville theory and the Schr\"{o}dinger equations, and use a version of the ODE/IM correspondence to write down a set of functional equations. By solving these functional equations, we can reproduce the well-known expression for the classical three-point function in the Liouville theory \cite{DO,ZZ}. We then provide a minimalistic review of the classical integrability of the string sigma model in AdS in \hyperref[lec3]{Lecture II} focusing on the roles of the monodromy and the quasi-momentum. After doing so, we set out to discuss our main topic; the three-point function in classical strings. As in the Liouville theory, one can convert the problem into a set of functional equations and their solutions provide nontrivial predictions for the classical three-point functions in string theory. We then speculate on how all these classical analyses could possibly be extended to the quantum level. 
\newpage

\section{Lecture I: Classical Three-point Functions in Liouville Theory\label{lec1}}
\begin{figure}
\centering
\includegraphics[clip,height=5cm]{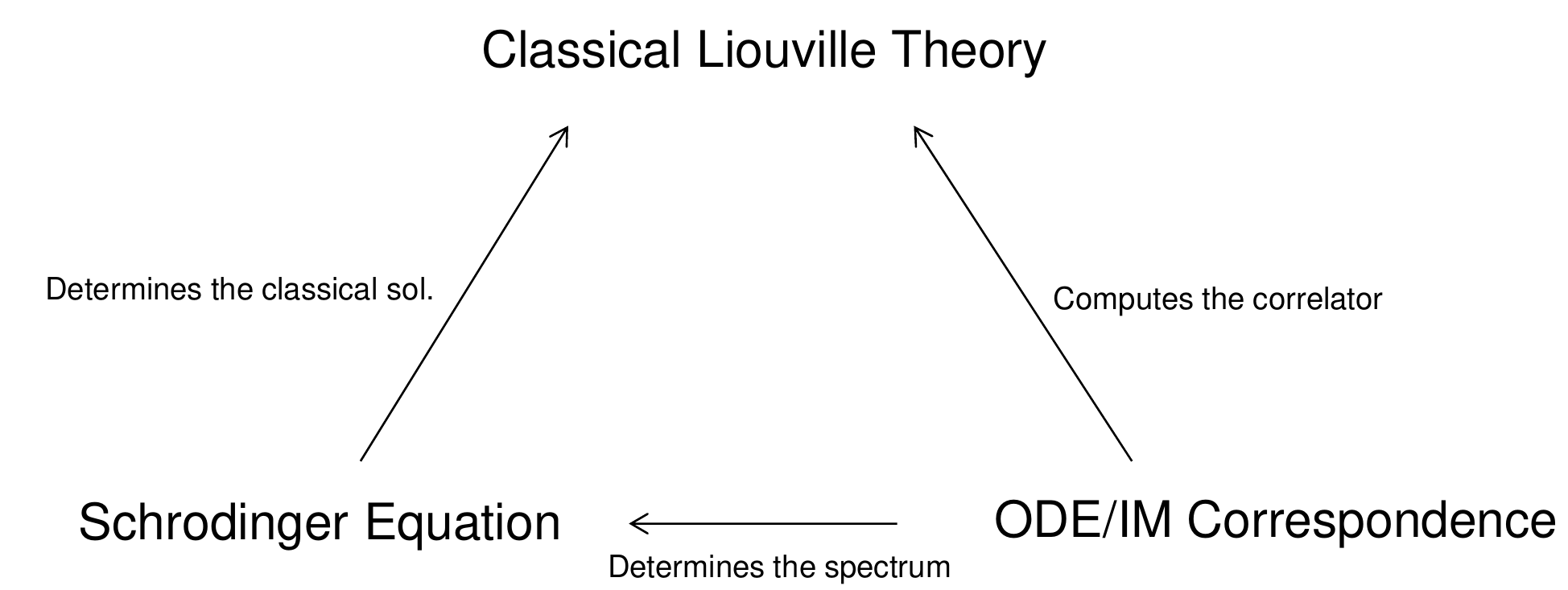}
\caption{The relations between the classical Liouville theory, the Schr\"{o}dinger equation and the ODE/IM correspondence.\label{fig1}}
\end{figure}

The main goal of three lectures is to learn how to use (a version of) the ODE/IM correspondence in the study of the AdS/CFT correspondence. In the first lecture, we explain that the classical Liouville theory plays an important role in the so-called ODE/IM correspondence as summarized in figure \ref{fig1}.
\subsection{Liouville field theory and its semi-classical limit}
Let us start by reviewing the basic properties of the Lioville field theory\footnote{A primary reference for the classical Liouville theory is \cite{Seiberg}.}. The Liouvillle field theory is a two-dimensional conformal field theory with continuous spectrum and is defined by the action
\begin{align}
S[\phi]=-\frac{1}{2\pi b^2} \int d^2 z \left( \partial \phi \bar{\partial} \phi+2 e^{2\phi}\right)\,.
\end{align}  
The parameter $b$ is the coupling constant and is related to the central charge as
\begin{align}
c=1+6\left(b+b^{-1} \right)^2\,.
\end{align}
The theory has a one-parameter family of local operators defined by
\begin{align}
\mathcal{V}_{\alpha}(z)\equiv e^{2\alpha \phi/b}\,,
\end{align}
with the conformal dimension $\Delta = \alpha (b+b^{-1}-\alpha)$. In the path integral formalism, the correlation function of the local operators is expressed as
\begin{align}\label{pathint}
\langle \mathcal{V}_{\alpha_1}(z_1)\cdots \mathcal{V}_{\alpha_n} (z_n)\rangle =\int \mathcal{D}\phi \left(\prod_{i=1}^{n}\mathcal{V}_{\alpha_i}(z_i)\right)e^{-S[\phi]}\,.
\end{align}

In the semi-classical limit, the coupling constant $b$ is sent to zero, $b\to 0$. In this limit, the action $S[\phi]$ diverges as $1/b^{2}$ and one can therefore approximate the path integral by its saddle point. If $\alpha_i$'s are $O(1)$, the presence of the operators $\mathcal{V}_{\alpha_i}$ does not change the saddle point since they scale as $\log \mathcal{V} \sim 1/b \ll 1/b^2$ and therefore are subleading. Those operators are called the light operators. On the other hand, if $\alpha_i$'s are $O(1/b)$, the contribution from the operators $\log  \mathcal{V}$ becomes comparable to the action and they produce source terms in the saddle-point equation. These operators do change the saddle point and are called the heavy operators\footnote{In terms of the conformal dimension and the central charge, the light operators are defined as the operators with $\Delta < O(\sqrt{c})$ whereas the heavy operators are defined as the operators with $\Delta = O(c)$ in the large $c$ limit.} in the literature. It is the heavy operator which is the subject of the rest of this lecture.

To discuss heavy operators, it is convenient to rewrite $\alpha_i$ as
\begin{align}
\alpha_i \equiv \eta_i /b\,.
\end{align}
Then the saddle-point approximation of the path integral \eqref{pathint} can be performed as
\begin{align}
\langle \mathcal{V}_{\alpha_1}(z_1)\cdots \mathcal{V}_{\alpha_n} (z_n)\rangle \overset{b\to 0}{\sim} \exp \left[ -\frac{\tilde{S}}{b^2}\right]\,,
\end{align}
with 
\begin{align}\label{defofStil}
\tilde{S}=\frac{1}{2\pi}\int d^2 z \left(\partial \phi_{\ast}\bar{\partial} \phi_{\ast}+2  e^{2\phi_{\ast}}-4\pi \sum_{i=1}^{n}\eta_i \phi_{\ast}\delta^{2}(z-z_i)\right)\,.
\end{align}
Here $\phi_{\ast}$ is the saddle-point value of $\phi$ satisfying the Liouville equation with source terms:
\begin{align}\label{lveq}
\partial \bar{\partial} \phi_{\ast}=e^{2\phi_{\ast}}-\pi \sum_{i=1}^{n}\eta_i \delta^2(z-z_i)\,.
\end{align}

\subsection{Classical action from the behavior around operators}
In principle, the semi-classical correlation function can be computed by solving the equation \eqref{lveq} and evaluating $\tilde{S}$.  There is however a trick which greatly simplifies the computation. For this purpose, let us first discuss the behavior of $\phi_{\ast}$ around the operator insertion points $z=z_i$. If we neglect the exponential term in \eqref{lveq}, one can easily determine the asymptotic behavior as
\begin{align}\label{phiasym}
\phi_{\ast}\overset{z\to z_i}{\sim} -2 \eta_i \log |z-z_i|+C_i +O(z-z_i)\,,
\end{align}
with $C_i$ being a constant independent of $z$. This approximation turns out to be valid as long as $\eta_i <1/2$ since in this region the exponential term is strictly smaller than $O(1/|z-z_i|^{2})$ whereas other terms in \eqref{lveq} scales as $O(1/|z-z_i|^{2})$ in the vicinity of $z_i$. In what follows we assume that $\eta_i$'s are in this parameter region.

We then consider the variation of $\tilde{S}$ with respect to $\eta_i$. There are in principle two different contributions to this variation: The first comes from the term proportional to $\eta_i$ in the definition of $\tilde{S}$ \eqref{defofStil}, whereas the second comes from the fact that the saddle-point solution $\phi_{\ast}$ itself depends on $\eta_i$ implicitly through the saddle-point equation \eqref{lveq}. However the second contribution is always of the form 
\begin{align}
\frac{\delta \tilde{S}}{\delta \phi_{\ast}}\frac{\delta \phi_{\ast}}{\delta \eta_i} \eta_i\,,
\end{align}
and therefore vanishes due to the saddle-point equation $\delta \tilde{S}/\delta \phi_{\ast}=0$. Thus we get a compact expression,
\begin{align}
\frac{\delta \tilde{S}}{\delta \eta_i}=-2 \phi_{\ast}(z_i)\,.
\end{align}
Using the asymptotic form of the solution \eqref{phiasym}, we can evaluate it as
\begin{align}\label{eq:discussiondiv}
\frac{\delta \tilde{S}}{\delta \eta_i}=\left.4 \eta_i \log |z-z_i|\right|_{z=z_i}-2 C_i\,.
\end{align}
The first term is obviously divergent and can be interpreted as the ``self energy" of the operator $\mathcal{V}_i$. This divergence can be cancelled unambiguosly by the renormalization of $\mathcal{V}_i$. Thus, after the renormalization, we  arrive at the formula
\begin{align}
\left. \frac{\delta \tilde{S}}{\delta \eta_i}\right|_{\rm renorm}=-2 C_i\,.
\end{align}
Upon using this formula, the computation of the semi-classical correlator boils down to the computation of constants $C_i$. In what follows, we will explain how we can compute $C_i$ by making use of a version of the ODE/IM correspondence.
\subsection{Relation to the Schr\"{o}dinger equation}
Let us first consider the stress-energy tensors of the Lioville theory. By the standard N\"{o}ether procedure, one can compute the stress-energy tensors as
\begin{align}
T(z)=-(\partial \phi_{\ast})^2 +\partial^2 \phi^{\ast}\,,\qquad \bar{T}(\bar{z})=-(\bar{\partial} \phi_{\ast})^2 +\bar{\partial}^2 \phi^{\ast}\,.
\end{align}
Although it may not so obvious at first sight, these quantities are holomorphic and anti-holomorphic respectively thanks to the Liouville equation \eqref{lveq}. Their form near the vertex operator can be determined from \eqref{phiasym} as
\begin{align}\label{Tasym}
T(z)\overset{z\to z_i}{\sim} \frac{\eta_i (1-\eta_i)}{(z-z_i)^2}\,,\qquad\qquad  \bar{T}(\bar{z})\overset{\bar{z}\to \bar{z}_i}{\sim} \frac{\eta_i (1-\eta_i)}{(\bar{z}-\bar{z}_i)^2}\,.
\end{align}
For three-point functions, the asymptotic behavior \eqref{Tasym} together with the condition that the stress energy tensors are not singular at infinity,
\begin{align}
T(z)\overset{z\to\infty}{\sim}O(1/z^{4})\,,\qquad \qquad \bar{T}(\bar{z})\overset{\bar{z}\to\infty}{\sim}O(1/\bar{z}^{4})\,,
\end{align}
completely determine\footnote{This is not the case for higher-point functions.} their forms as\footnote{There is also a corresponding formula for $\bar{T}(\bar{z})$ but we omit writing it since it is just a complex conjugation of \eqref{t3explicit}.}
\begin{align}\label{t3explicit}
\begin{aligned}
T(z)=\frac{1}{(z-z_1)(z-z_2)(z-z_3)}\sum_{i\neq j\neq k}\frac{\eta_i (1-\eta_i)z_{ij}z_{ik}}{z-z_i}
\end{aligned}
\end{align}
with $z_{ij}\equiv z_i-z_j$.

An important property of the classical Liouville equation is that one can reconstruct the solution from these stress energy tensors:
\begin{itemize}
\item[] \textbf{Reconstruction formula}\footnote{One can easily derive this formula by using the fact that $e^{-\phi}$ satisfies both the holomorphic and anti-holomorphic Schr\"{o}dinger equation and there are only two linearly independent solutions to each equation. The normalization condition can be derived by plugging in the expression \eqref{reconst} into the Liouville equation.}\\
The classical solution to the Liouville equation can be expressed as
\begin{align}\label{reconst}
e^{-\phi_{\ast}}=\psi_1(z)\tilde{\psi}_1(\bar{z}) -\psi_2(z) \tilde{\psi}_2(\bar{z}) \,,
\end{align}
where $\psi_i$ and $\bar{\psi}_i$ are the solutions to the Schr\"{o}dinger equations,
\begin{align}\label{Schr}
\left(\partial^2 +T(z)\right) \psi_i (z)=0\,,\qquad \left(\bar{\partial}^2 +\bar{T}(\bar{z})\right) \tilde{\psi}_i (\bar{z})=0\,
\end{align}
satisfying the normalization condition
\begin{align}
\langle \psi_1 ,\psi_2 \rangle \langle \tilde{\psi}_1,\tilde{\psi}_2\rangle =1 \,.
\end{align}
\end{itemize}
Here $\langle A,B\rangle$ is called the Wronskian and is defined by $\langle A,B \rangle \equiv A\partial B-B\partial A$ or $\langle A,B \rangle \equiv A\bar{\partial} B-B\bar{\partial} A$. A crucial property of the Wronskians is that it is independent of the coordinates $z$ and $\bar{z}$ if two entries ($A$ and $B$) satisfy the same Schr\"{o}dinger equation.

Now comes a word of caution: Although the expression \eqref{reconst} always provides a formal solution to the Lioville equation, it is not necessarily single-valued on the Riemann surface. To have a single-valued solution, we need to choose $\psi_i$ and $\tilde{\psi}_i$ appropriately. Let us show how this can be done for the three-point function. From the asymptotic form of the stress energy tensors, we can compute two independent solutions $i_{\pm}$ around each $z_i$:
\begin{align}
\begin{aligned}
&i_{+}\sim \frac{(z-z_i)^{\eta_i}}{\sqrt{1-2\eta_i}}\,,\qquad\qquad&& i_{-}\sim \frac{(z-z_i)^{1-\eta_i}}{\sqrt{1-2\eta_i}}\,,\\
&\bar{i}_{+}\sim \frac{(\bar{z}-\bar{z}_i)^{\eta_i}}{\sqrt{1-2\eta_i}}\,,\qquad\qquad&& \bar{i}_{-}\sim \frac{(\bar{z}-\bar{z}_i)^{1-\eta_i}}{\sqrt{1-2\eta_i}}\,,
\end{aligned}
\end{align}
The factors $1/\sqrt{1-2\eta_i}$ are needed for the normalization we imposed $\langle i_{+},i_{-} \rangle=1$.

Using this basis of solutions, one can construct a single-valued solution around $z_1$ as
\begin{align}\label{singlez1}
e^{-\phi_{\ast}}=A 1_{+}\bar{1}_{+}-A^{-1}1_{-}\bar{1}_{-}\,,
\end{align}
where $A$ is a constant which cannot be fixed purely from the single-valuedness around $z_1$. In terms of $A$, the constant $C_1$ can be expressed as
\begin{align}\label{C1A}
C_1 =\log (1-2\eta_1) -\log A\,.
\end{align}
So the remaining task is to determine $A$ by imposing the single-valuedness around other operators. For this purpose, we change the basis  of solutions using
\begin{align}\label{basistransf}
1_{\pm} =\langle 1_{\pm} ,2_{-}\rangle 2_{+}-\langle 1_{\pm},2_{+}\rangle 2_{-}\,.
\end{align}
These relations can be easily confirmed by computing the Wronskians between $2_{\pm}$ and the left or the right hand side of \eqref{basistransf}, and using the normalization condition $\langle i_{+},i_{-} \rangle=1$. After the change of basis, \eqref{singlez1} reads
\begin{align}\label{singlez2}
\begin{aligned}
e^{-\phi_{\ast}}=&\left(\cdots \right)2_{+}\bar{2}_{+} +\left(\cdots \right)2_{-}\bar{2}_{-}\,\\
&-\left(A \langle 1_{+},2_{-}\rangle\langle \bar{1}_{+},\bar{2}_{+}\rangle-A^{-1}\langle 1_{+},2_{-}\rangle\langle \bar{1}_{-},\bar{2}_{+}\rangle \right)2_{+}\bar{2}_{-}\\
&-\left(A \langle 1_{+},2_{+}\rangle\langle \bar{1}_{+},\bar{2}_{-}\rangle-A^{-1}\langle 1_{-},2_{+}\rangle\langle \bar{1}_{-},\bar{2}_{-}\rangle \right)2_{-}\bar{2}_{+}\,.
\end{aligned}
\end{align}
In \eqref{singlez2}, the terms in the first line are single-valued also around $z_2$ whereas the terms on the last two lines violate the single-valuedness. Therefore, we need to impose that the coefficients of these two last terms must be zero. This allows us to determine $A$ unambiguously as
\begin{align}\label{Aexp}
A=\sqrt{\frac{\langle1_{-},2_{-} \rangle\langle\bar{1}_{-},\bar{2}_{+} \rangle}{\langle1_{+},2_{-} \rangle\langle\bar{1}_{+},\bar{2}_{+} \rangle}}\,.
\end{align}
Combining \eqref{C1A} and \eqref{Aexp}, one can reduce the problem of computing $C_i$ to the computation of the Wronskians $\langle i_{\pm} ,j_{\pm}\rangle $.

Before ending this subsection, let us make one additional remark. For the case of the three-point function, the single-valuedness around $z_1$ and $z_2$ automatically guarantees the single-valuedness around $z_3$. This follows from the fact that the path going around both $z_1$ and $z_2$ can be continuously deformed into the path around $z_3$. This is of course only true for the case of three-point function and the analysis will become more complicated for the higher-point functions.
\subsection{Wronskians and functional equations\label{subsec:wronsfunc}}
To compute the Wronskians, we  proceed in three steps. The first step is to introduce an additional complex parameter to our story which we call the ``spectral parameter", appealing to the analogy with the classical integrable system.  The idea is to replace $\eta_i$'s with the following quantity by introducing a new parameter $\theta$:
\begin{align}\label{eq:repl}
\eta_i \to \frac{1}{2}-\eta_i e^{\theta}\,.
\end{align} 
This replacement may seem a bit artificial at first sight, but it is actually motivated by the structure of the stress energy tensor, which before the replacement reads
\begin{align}
T(z)\overset{z\to z_i}{\sim} \frac{\eta_i (1-\eta_i)}{(z-z_i)^2}\,.
\end{align}
This expression is invariant under the exchange of $\eta_i$ and $1-\eta_i$ and it translates to the invariance under the simple transformation $\theta \to \theta-\pi i$, after the replacement \eqref{eq:repl}. Now, after the replacement, the solutions to the Schr\"{o}dinger equation around $z=z_i$ reads
\begin{align}
\begin{aligned}\label{itheta}
i_{\pm}(\theta)\overset{z\to z_i}{\sim} \frac{(z-z_i)^{\frac{1}{2}\mp \eta_i e^{\theta}}}{\sqrt{2\eta_i e^{\theta}}} 
\end{aligned}
\end{align}
One important advantage of having the spectral parameter is that two solutions swap their roles if we shift the spectral parameter by $\pm i \pi$. This allows us to choose the normalization of the solutions such that they satisfy
\begin{align}
i_{+} (\theta- \pi i)= i_{-}(\theta)\,,\qquad i_{-}(\theta-i\pi)=-i_{+}(\theta)\,.
\end{align}
In the notation common in the AdS/CFT integrability, $f^{\pm} (x)\equiv f(x\pm i\pi/2)$, they read $i_{+}^{--}=i_{-}$ and $i_{-}^{--}=-i_{+}$.

\begin{figure}
\centering
\includegraphics[clip, height=5cm]{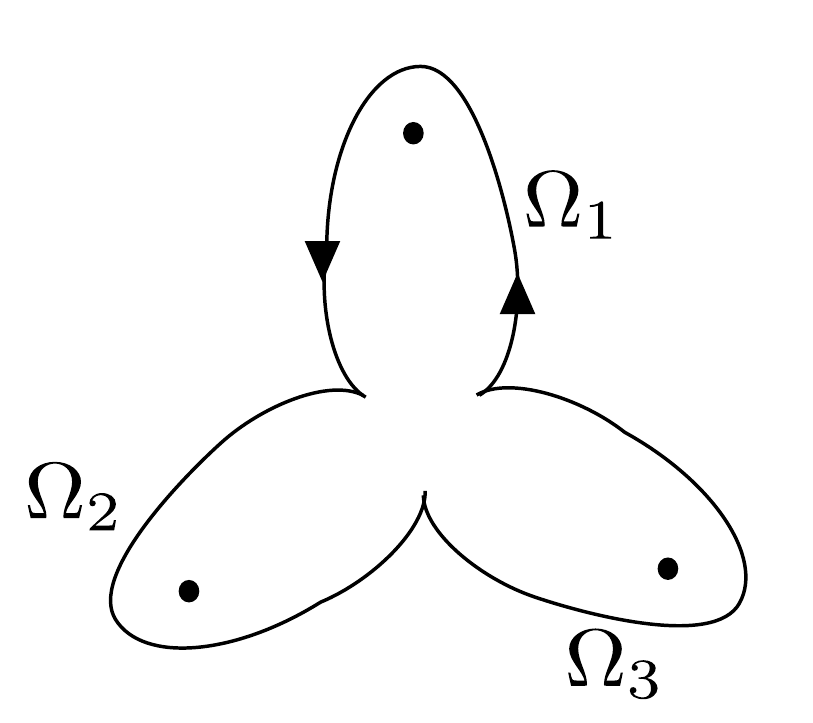}
\caption{Monodromy relation for the three-point function. The three-point function is described by a thrice-punctured sphere. Around each puncture, one can define a monodromy matrix $\Omega_i$. The multiplication of three monodromy matrices $\Omega_1\Omega_2\Omega_3$ corresponds to the monodromy around the contour depicted above, which can be shrunk to a point. They therefore satisfy $\Omega_1\Omega_2\Omega_3={\bf 1}$.\label{fig2}}
\end{figure}

The next step is to consider the ``monodromy". Solutions to the Schr\"{o}dinger equation \eqref{Schr} (or more precisely the spectral-parameter deformation of it) are not single-valued on the complex plane. For instance, owing to the behavior \eqref{itheta}, the solutions $1_{\pm}$ acquire  phases when they go around $z_1$.
\begin{align}\label{monod1}
1_{+} \to e^{ i p_1 (\theta)} 1_{+} \,, \qquad 1_{-} \to e^{ i( -p_1(\theta))} 1_{-}\,.
\end{align}
with $p_i$ given by
\begin{align}
p_i (\theta)=2\pi \left( \frac{1}{2}-\eta_i e^{\theta}\right)
\end{align}
If they instead go around $z_2$ or $z_3$, the resulting solution will be linear combination of $1_{+}$ and $1_{-}$ and it will be summarized as the action of a $2 \times 2$ matrix $\Omega_i$,
\begin{align}
\left(\begin{array}{c}1_{+}\\1_{-}\end{array}  \right)\to \Omega_i\left(\begin{array}{c}1_{+}\\1_{-}\end{array}  \right)\,,
\end{align}
where the subscript $i$ specifies which point they go around. An important property of these monodromies is the fact that the product of three monodromies must be trivial (see figure \ref{fig2}):
\begin{align}\label{monodrela}
\Omega_1 \Omega_2 \Omega_3 ={\bf 1}\,.
\end{align} 
This is because the contour which goes around three points $z_1$, $z_2$ and $z_3$ is equivalent to the contour which does not encircle any point.

In this monodromy representation, the transformation property \eqref{monod1} translates to
\begin{align}
\Omega_1 = \left(\begin{array}{cc}e^{i p_1}&0 \\0&e^{-  i p_1} \end{array} \right)\,.
\end{align}
On the other hand, $\Omega_2$ is not diagonal in general. However the entries of the matrix,
\begin{align}
\Omega_2 \equiv \left(\begin{array}{cc}a&b \\c&d \end{array}\right)
\end{align}
 are constrained by the fact that the eigenvalues of $\Omega_2$ are given by $e^{ip_2}$ and $e^{-ip_2}$:
\begin{align}
&a+d = 2 \cos p_2\,,\label{one}\\
&ad-bc=1\,.\label{two}
\end{align}
Similarly, $\Omega_3$ is not diagonal either. Using the relation \eqref{monodrela}, one can express its entries as
\begin{align}
\Omega_3 = \Omega_2^{-1}\Omega_1^{-1}=\left( \begin{array}{cc}e^{-ip_1}d&-e^{ip_1}b\\-e^{-ip_1}c&e^{ip_1}a\end{array}\right)\,.
\end{align}
Since its eigenvalues are given by $e^{\pm i p_3}$, one gets another constraint
\begin{align}
e^{-ip_1}d+e^{ip_1}a=2 \cos p_3\,.\label{three}
\end{align}
Combining three constraints \eqref{one}, \eqref{two} and \eqref{three}, one can determine $a$, $d$ and $bc$. This allows us to write down explicitly the entries of $\Omega_2$ and $\Omega_3$ up to one unknown constant. One can then diagonalize these matrices to read off how $2_{\pm}$ and $3_{\pm}$ can be expressed in terms of $1_{\pm}$ carrying out these simple but tedious analyses, we arrive at the following expressions for the product of Wronskians:
\begin{align}
\langle i_{+},j_{+}\rangle\langle i_{-},j_{-}\rangle&=\frac{\sin \left(\frac{p_i+p_j+p_k}{2} \right)\sin \left(\frac{p_i+p_j-p_k}{2} \right)}{\sin p_i \sin p_j}\,.
\end{align}
Now using the relation $i_{\pm}^{--}\sim i_{\mp}$, we can convert it into a functional equation involving only $\langle i_{+},j_{+}\rangle$,
\begin{align}
\langle i_{+},j_{+}\rangle\langle i_{+},j_{+}\rangle^{--}&=\frac{\sin \left(\frac{p_i+p_j+p_k}{2} \right)\sin \left(\frac{p_i+p_j-p_k}{2} \right)}{\sin p_i \sin p_j}\,.
\end{align}
This equation takes a form similar to the {\it crossing equation} for the scalar factor of the integrable S-matrix\footnote{See for instance a part of the big review on the AdS/CFT integrability \cite{dressing}.}.
\subsection{Solving the functional equation}
A solution to this functional equation can be constructed by making use of the property of $p_i$
\begin{align}
p_i^{--}=2\pi -p_i\,,
\end{align} 
and the identity
\begin{align}
\sin x =\frac{\pi}{\Gamma (x/\pi)\Gamma (1- x/\pi)}\,.
\end{align}
It is given explicitly by
\begin{align}\label{ipm}
\begin{aligned}
\langle i_{+},j_{+}\rangle&=\frac{\textcolor[rgb]{0,0,1}{e^{f(\theta)}}}{\sqrt{(1-2p_i)(1-2p_j)}} \frac{\Gamma (p_i/\pi)\Gamma (p_j/\pi)}{\Gamma \left(\frac{p_i+p_j+p_k}{2\pi}-1\right)\Gamma \left(\frac{p_i+p_j-p_k}{2\pi}\right)}\,,\\
\langle i_{-},j_{-}\rangle&=\frac{\textcolor[rgb]{0,0,1}{e^{-f(\theta)}}}{\sqrt{(1-2p_i)(1-2p_j)}} \frac{\Gamma (2- p_i/\pi)\Gamma (2-p_j/\pi)}{\Gamma \left(1-\frac{p_i+p_j+p_k}{2\pi}\right)\Gamma \left(2-\frac{p_i+p_j-p_k}{2\pi}\right)}\,,
\end{aligned}
\end{align}
where $f(\theta)$ can be any periodic function in $\theta$, $f^{--}=f$, and is the analogue of the so-called CDD factor \cite{CDD}, which represents the ambiguity of the solution to the functional equation.

To determine the CDD ambiguity, we need some extra input on the functions $\langle i_{+},j_{+}\rangle$. In the present case, it comes from the WKB expansion of the Schr\"{o}dinger equation: As shown below, in the spectral-parameter deformed Schr\"{o}dinger equation, $e^{\theta}$ plays the role of $1/\hbar$ in the standard Schr\"{o}dinger equation,
\begin{align}
\left(\partial^2 + T(z;\theta)\right)\psi\quad \overset{\theta\to\infty}{\sim}\quad \left(\partial^2-e^{2\theta}t(z)\right)\psi=0\,,
\end{align}
with
\begin{align}
t(z)=\frac{1}{(z-z_1)(z-z_2)(z-z_3)}\sum_{i\neq j\neq k}\frac{\eta_i^2 z_{ij}z_{ik}}{z-z_i}\,.
\end{align}
 We can therefore apply the WKB expansion to study the large positive $\theta$ limit. For instance, the WKB solutions near $z_i$ are given by
\begin{align}\label{wkbsol}
\psi \sim\lim_{\epsilon\to 0}\frac{\textcolor[rgb]{0,0,1}{\epsilon^{\pm \eta_i e^{\theta}-\frac{1}{2}}}}{\sqrt{2t(z)e^{\theta}}}\exp \left( \pm e^{\theta}\int_{z_i+\epsilon}^{z} \sqrt{t(z)}\,dz\right)\,,
\end{align}
where the factor denoted in blue was introduced in order to match it with the asymptotic form of $i_{\pm}$ \eqref{itheta}. The expression \eqref{wkbsol} gives two formal solutions depending on the choice of the sign in the exponent. However, among these two solutions, the one which is larger around $z_i$ is ambiguous since one can always add to it the other solution, which is exponentially small, without changing the leading exponential behavior. This is one of the important properties of the WKB expansion although it may not be emphasized much in other contexts. In the present case, where $\theta$ is large and positive, the one with a plus sign in the exponent in \eqref{wkbsol} is a small\footnote{Throughout this section, we assume that $\eta_i$'s are positive and real.} solution and corresponds to $i_{-}$. Therefore, in this regime of $\theta$, one can reliably use the WKB expansion only for the Wronskians which involve two $i_{-}$'s; namely $\langle i_{-},j_{-}\rangle$. It is a straightforward exercise\footnote{See \cite{HK} for details.} to compute this Wronskian from \eqref{wkbsol}. By comparing it with the expression \eqref{ipm}, we can determine $f(\theta)$ as
\begin{align}
f(\theta) = -\left( \eta_i \log \frac{z_{ij}z_{ik}}{z_{jk}}+\eta_j \frac{ z_{ij}z_{jk}}{z_{ki}}\right)\,.
\end{align} 
Note also that $\log \langle i_{-},j_{-} \rangle$ in \eqref{ipm} does not have any singularities\footnote{In other words, $\langle i_{-},j_{-}\rangle$ does not have any poles or zeros.} when $\theta$ is real and positive. This is consistent with the fact that $\langle i_{-},j_{-} \rangle$ can be consistently expanded into the WKB series in this parameter range\footnote{When a function can be expanded safely into a WKB series, we do not expect any singularities in the expression. However, it is difficult to make such expectations into a more rigorous argument.}.

Having computed the Wronskian, we can now backtrack all the steps and compute the semi-classical three-point function. The final result reads
\begin{align}
\langle \mathcal{V}_1\mathcal{V}_2\mathcal{V}_3\rangle \sim \frac{C_{123}}{|z_{12}|^{\Delta_1+\Delta_2-\Delta_3}|z_{23}|^{\Delta_2+\Delta_3-\Delta_1}|z_{31}|^{\Delta_3+\Delta_1-\Delta_2}}
\end{align}
where $\Delta_i = \eta_i (1-\eta_i)/b^{2}$ and $C_{123}$ is given by
\begin{align}
\begin{aligned}\label{finalliouville}
b^{2}\log C_{123}\sim \sum_{i} F(2\eta_i)-F(\eta_1+\eta_2+\eta_3)-\sum_{i\neq j\neq k}F(\eta_i+\eta_j-\eta_k)\,,
\end{aligned}
\end{align}
with
\begin{align}
F(\eta )\equiv \int^{\eta}_{1/2}\log \left( \frac{\Gamma(x)}{\Gamma(1-x)}\right)dx\,.
\end{align}
This matches with the classical limit of the DOZZ formula \cite{DO,ZZ}.
\newpage
\section{Lecture II: Classical Three-point Functions for Strings in AdS$_2$\label{lec3}}
In the second part of the lecture, we study the classical three-point functions of string in AdS by applying the ideas explained in \hyperref[lec1]{Lecture I}. 

The AdS$_5$/CFT$_4$ correspondence relates the planar $\mathcal{N}=4$ supersymmetric Yang-Mills (SYM) theory to string theory on $AdS_5\times S^{5}$ spacetime. In particular, the strong coupling limit of the three-point function in $\mathcal{N}=4$ SYM can be identified with the classical three-point function of the string sigma model. In what follows, we first  briefly review the integrability of the string sigma model and then explain how the ideas in the previous lecture can be applied to the string theory.
\subsection{Classical strings in AdS$_2$ and integrability}
Here we give a quick review of the classical strings in AdS$_2$ and its integrability (see also \cite{JW}).
\subsubsection*{Action and equation of motion}
In this lecture, we focus on the string moving in the AdS$_2$ subspace of the full $AdS_5\times S^5$ spacetime. AdS$_2$ is defined by the embedding coordinate as
\begin{align}
\eta_{\mu\nu}X^{\mu}X^{\nu} =-1\,, \qquad \mu,\nu=(-1,1,4)\,,\qquad \eta_{\mu\nu}=(-1,1,1)\,.
\end{align}
To study the correlation functions, it is convenient to experss them using the so-called Poincare coordinates
\begin{align}\label{defPoinCare}
\begin{aligned}
&X^{-1}+X^{4}=\frac{1}{{\sf z}}\,,\qquad X^{-1}-X^{4}={\sf z}+\frac{{\sf x}^2}{{\sf z}} \,,\qquad X^{1}=\frac{{\sf x}}{{\sf z}}\,.
\end{aligned}
\end{align}
In terms of the Poincare coordniates the boundary of the AdS spacetime is at ${\sf z}=0$.

The Polyakov action of the string sigma model on AdS$_2$ reads
\begin{align}\label{polac}
S=\frac{\sqrt{\lambda}}{\pi} \int d^2z \left[\partial X^{\mu}\bar{\partial} X_{\mu}+\Lambda (X^{\mu}X_{\mu}+1)\right]\,,
\end{align}
where $\Lambda$ is the Lagrange multiplier. By taking the variation of this action and eliminating $\Lambda$, we arrive at the following equation of motion
\begin{align}\label{eomAdS}
\partial \bar{\partial} X^{\mu}- (\partial X^{\nu}\bar{\partial}X_{\nu})X^{\mu}=0\,.
\end{align}
Because of the nonlinearity of the equation, it is not straightforward to solve this equation. To make progress, it is useful to rewrite it into the following equation:
\begin{align}\label{laxAdS}
\left[\partial + \frac{j_{z}}{1-x},\quad \bar{\partial}+\frac{j_{\bar{z}}}{1+x}\right]=0\,,
\end{align}
with
\begin{align}
\begin{aligned}\label{jandg}
j\equiv g^{-1}dg \,,\qquad g=\left(\begin{array}{cc}X^{-1}+X^{4}&X^{1}\\X^{1}&X^{-1}-X^{4}\end{array} \right)\,.
\end{aligned}
\end{align}
The parameter $x$ in \eqref{laxAdS} is called the spectral parameter and one can show that \eqref{laxAdS} is satisfied for any $x$ as long as $X^{\mu}$'s satisfy the equation of motion \eqref{eomAdS}.
\subsubsection*{Linear problem}
An advantage of this reformulation is that it makes manifest the classical integrability of the theory. Owing to the flatness of the connection \eqref{laxAdS}, the following quantity does not depend on small deformations of the contour $C$:
\begin{align}
t(x)\equiv {\rm Tr}\left[ {\rm P}\exp \left(-\oint_{C} \frac{j_z dz}{1-x}+\frac{j_{\bar{z}}d\bar{z}}{1+x}\right)\right]\,.
\end{align} 
This means that $t(x)$ is a conserved quantity. By expanding it in $x$, one can get infinitely many conserved charges as coefficients of the expansion. 

The flatness condition \eqref{laxAdS} allows us to consider an associated linear problem
\begin{align}\label{linAdS}
\left(\partial + \frac{j_z}{1-x}\right) \psi=0\,, \qquad \left(\bar{\partial}+\frac{j_{\bar{z}}}{1+x} \right)\psi=0\,,
\end{align}
where $\psi$ is a two-component vector defined on the world-sheet. These linear equations are the analogue of the Schr\"{o}dinger equation for the Liouville theory, and one can similarly reconstruct the solution to the equation of motion from the solutions to this linear problem as we see later.
\subsubsection*{Simple solutions}
Let us now consider solutions to the equation of motion \eqref{eomAdS}. The simplest solution is given by
\begin{align}
{\sf z}=e^{2\kappa\tau} \,,\qquad {\sf x}=0\,.
\end{align}
where $\tau$ is the Euclidean world-sheet time defined by $z=e^{\tau+i\sigma}$. This solution describes a string emitted at the origin of the boundary of AdS and absorbed by the horizon, and corresponds to a two-point function $\langle \mathcal{O}(0)\mathcal{O}(\infty)\rangle$ of the scalar operators with the dimension\footnote{The relation between $\kappa$ and $\Delta$ can be determined by computing the N\"{o}ether charge
\begin{align}
\Delta= \sqrt{\lambda}{\pi}\int^{2\pi}_{0}d\sigma \frac{\partial_{\tau}{\sf z}}{2{\sf z}}=2\sqrt{\lambda}\kappa\,. 
\end{align}
} $\Delta = 2\sqrt{\lambda} \kappa$. For this solution, the linear problem also takes a very simple form
\begin{align}
\left(\partial -\frac{\kappa}{(1-x)z}\left( \begin{array}{cc}1&0\\0&-1\end{array}\right) \right)\psi=0\,,\qquad \left(\bar{\partial} -\frac{\kappa}{(1+x)\bar{z}}\left( \begin{array}{cc}1&0\\0&-1\end{array}\right) \right)\psi=0\,,
\end{align}
and the solutions read
\begin{align}
\psi_{+}=\left( \begin{array}{c}0\\z^{\frac{-\kappa}{1-x}}\bar{z}^{\frac{-\kappa}{1+x}}\end{array}\right) \,,\qquad \psi_{-}=-\left( \begin{array}{c}z^{\frac{\kappa}{1-x}}\bar{z}^{\frac{\kappa}{1+x}}\\0\end{array}\right)\,.
\end{align}
Here we normalized the two solutions so that they satisfy
\begin{align}\label{eq:psiplusminuswron}
\langle \psi_{+}\,,\psi_{-}\rangle\equiv \epsilon_{ab}\psi_{+}^{a}\psi_{-}^{b}=1\,.
\end{align}
The bracket $\langle A,B \rangle$ is the counterpart of the Wronskian for the Schr\"{o}dinger equation and is independent of $z$ and $\bar{z}$ if $A$ and $B$ satisfy the linear equations \eqref{linAdS}.
As in the case of the Schr\"{o}dinger equation, these solutions have monodromy around $z=0$,
\begin{align}
\psi_{\pm} \to e^{\pm ip(x)}\psi_{\pm}
\end{align}
where $p$ is called the quasi-momentum and is given by
\begin{align}
p(x)=\frac{2\pi\Delta x}{\sqrt{\lambda}(x^2-1)}\,.
\end{align}

A more general solution which describes the string emitted from the boundary at ${\sf x}={\sf x}_0$ can be obtained from the above solution by performing the translation:
\begin{align}\label{simplesol2}
{\sf z}=e^{2\kappa\tau} \,,\qquad {\sf x}={\sf x}_0\,.
\end{align}
Then the solutions to the linear problem become
\begin{align}
\psi_{+}=z^{\frac{-\kappa}{1-x}}\bar{z}^{\frac{-\kappa}{1+x}}\left(\begin{array}{c}-{\sf x}_0\\1\end{array} \right)\,,\qquad \psi_{-}=-\frac{z^{\frac{\kappa}{1-x}}\bar{z}^{\frac{\kappa}{1+x}}}{1+{\sf x}_0^2}\left(\begin{array}{c}1\\{\sf x}_0\end{array} \right)\,.
\end{align}
Note that the monodromy property of these solutions are the same as before, $\psi_{\pm}\to e^{\pm ip}\psi_{\pm}$. 
\subsubsection*{$Z_2$-symmetry}
The string sigma model in AdS$_2$ has an interesting $Z_2$-symmetry. From the structure of $g$ given in \eqref{jandg}, one can easily verify that $g$ satisfies
\begin{align}
\sigma_2 \,g\, \sigma_2 =g^{-1}\,,
\end{align}
where $\sigma_2$ is the standard Pauli matrix. Using this property, one can show that the current $g^{-1}dg$ flips the sign under the conjugation by $\sigma_2 g$ as
\begin{align}
g^{-1}\sigma_2 (g^{-1}dg)\sigma_2 g =-g^{-1}dg\,,
\end{align}
and the connection \eqref{laxAdS} transforms as
\begin{align}
\begin{aligned}
&g^{-1}\sigma_2\left(\partial + \frac{j_z}{1-x} \right)\sigma_2 g=\partial - \frac{j_z}{1-x}+j_z=\partial + \frac{j_z}{1-1/x}\,,\\
&g^{-1}\sigma_2\left(\bar{\partial} + \frac{j_{\bar{z}}}{1+x} \right)\sigma_2 g=\bar{\partial} - \frac{j_{\bar{z}}}{1+x}+j_{\bar{z}}=\bar{\partial} + \frac{j_{\bar{z}}}{1+1/x}\,.
\end{aligned}
\end{align}
We can therefore conclude that the linear problems with the spectral parameter $x$ and $1/x$ are related by the conjugation $\sigma_2 g$. Since the quasi-momentum satisfies $p(1/x)=-p(x)$, the solutions to the linear problem $\psi_{\pm}$ tranform as
\begin{align}\label{eq:Z2symmetrypsi}
g^{-1}\sigma_2 \psi_{+} (x)= i\psi_{-}(1/x) \,,\qquad g^{-1}\sigma_2 \psi_{-} (x)=i \psi_{+}(1/x)\,.
\end{align} 
Note that the extra factors of $i$ are necessary to preserve the normalization condition \eqref{eq:psiplusminuswron}.
To make clear the analogy with the Liouville theory, it is useful to parametrize the spectral parameter in a different way as
\begin{align}
e^{\theta}=\frac{x+1}{x-1}\,.
\end{align} 
In this parametrization, the flat connection becomes
\begin{align}
\left[ \partial + \frac{1-e^{\theta}}{2}j_z,\quad \bar{\partial}+\frac{1-e^{-\theta}}{2}j_{\bar{z}}\right]=0\,,
\end{align}
and the $Z_2$-symmetry reads
\begin{align}
g^{-1}\sigma_2 \psi_{+}= i\psi_{-}^{--} \,,\qquad g^{-1}\sigma_2 \psi_{-}=i\psi_{+}^{--}\,,
\end{align}
where as before $f^{--}$ denotes the shift of the argument by $-\pi i$. 
\subsubsection*{Reconstruction formula}
So far, we discussed how to obtain linear problems from the original non-linear equation of motion of the sigma model. It turns out that one can also go in the other direction: Namely, one can construct the solution to the non-linear equation of motion from the solutions to the linear problems as follows:
\begin{align}\label{reconAdS}
g^{-1}=\left.(\psi_1,\psi_2)\right|_{x=0}\,.
\end{align} 
Here $\psi_{1,2}$ are two linearly independent solutions to the equation of motion which satisfy the normalization condition
\begin{align}
\langle \psi_1\,,\psi_2\rangle \equiv \epsilon_{ab}\psi_1^{a}\psi_2^{b}=1\,.
\end{align}

The relation \eqref{reconAdS} follows from the fact that $g^{-1}$ satisfies
\begin{align}
\left(\partial +j_z\right)g^{-1}=\left(\bar{\partial}+j_{\bar{z}}\right)g^{-1}=0\,,
\end{align}
while the normalization condition is needed to guarantee $\det g =1$.
\subsection{Classical limit and vertex operators}
Let us now discuss the three-point function. The AdS/CFT correspondence relates the three-point function in planar $\mathcal{N}=4$ SYM and the three-point function on the string worldsheet:
\begin{align}
\langle\mathcal{O}_{\Delta_1}({\sf x}_1)\mathcal{O}_{\Delta_2}({\sf x}_2)\mathcal{O}_{\Delta_3}({\sf x}_3) \rangle_{\text{$\mathcal{N}=4$ SYM}} = \int \mathcal{D}X\, \mathcal{V}_1 (z_1)\mathcal{V}_2 (z_2)\mathcal{V}_3 (z_3) e^{-S[X]} \,.
\end{align}
Here the right hand side is the path integral of the string sigma model. Both the vertex operators and the action consist of the AdS$_5$ part and the S$^{5}$ part\footnote{There are of course contributions from fermionic degrees of freedom. However, they can be ignored in the classical limit.}. However, in this lecture we focus on the AdS part. In addition, we only discuss {\it scalar} operators in $\mathcal{N}=4$ SYM and assume that they live on a one-dimensional subspace, which is dual to AdS$_2$. In such cases, (the AdS-part of) the vertex operator takes a simple form:
\begin{align}
\mathcal{V}_i(z_i)=\left({\sf z}+\frac{({\sf x}-{\sf x}_i)^2}{{\sf z}} \right)^{-\Delta_i}(z_i)={\rm Tr}\left[\left(\begin{array}{cc}1&{\sf x}_i\\{\sf x}_i&{\sf x}_i^{2}\end{array} \right)g^{-1}\right]^{-\Delta_i}(z_i)
\end{align}

As in the previous lecture, this path integral can be approximated by its saddle point when the coupling constant $\sqrt{\lambda}$ is very large\footnote{Recall the form of the action \eqref{polac}.}. When the operators have dimensions $\Delta = O(\sqrt{\lambda})$, the vertex operators cannot be neglected when determining the saddle point. In the presence of such ``heavy" operators, the saddle-point solution takes a form of a three-pronged sphere. To compute the classical three-point function, we utilize a trick similar to the one we employed for the Liouville theory. Namely, instead of directly solving the saddle-point equation and evaluating the action, we consider the variation of the three-point function under the change of $\Delta_i$. As we discussed in the previous lecture, non-vanishing contribution comes only from the explicit dependence on $\Delta_i$ contained in the vertex operators. This leads to a formula for the AdS contribution of the classical three-point function,

\begin{align}\label{eq:contributionfromvertex}
\frac{\partial \log \langle\mathcal{O}_{\Delta_1}\mathcal{O}_{\Delta_2}\mathcal{O}_{\Delta_3} \rangle }{\partial \Delta_i}=-\log {\rm Tr}\left[\left(\begin{array}{cc}1&{\sf x}_i\\{\sf x}_i&{\sf x}_i^{2}\end{array} \right)g^{-1}\right] (z_i)\,.
\end{align}
This formula allows us to compute the classical three-point function from the behavior of $g$ around vertex operators. In what follows, we re-express this formula in terms of the Wronskians of the solutions to the linear problem.
\subsection{Vertex operators from Wronskians}
\begin{figure}
\centering
\includegraphics[clip,height=4cm]{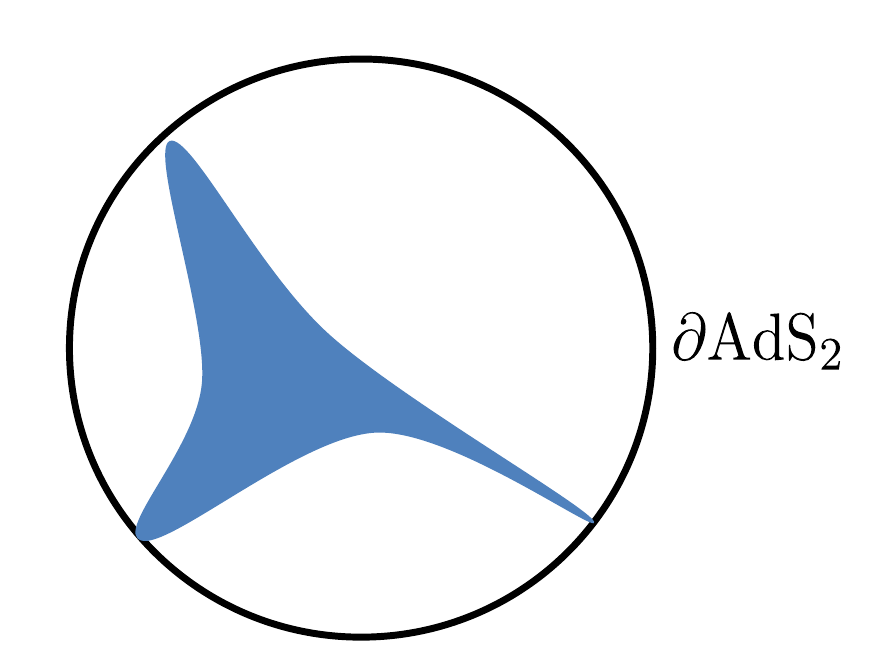}
\caption{The target space picture for the classical three-point function of AdS$_2$ string. In the target space, it corresponds to a surface (denoted in blue) with three legs where each leg is anchored at the operator insertion point on the AdS$_2$ boundary.\label{fig3}}
\end{figure}

To evaluate the right hand side of \eqref{eq:contributionfromvertex}, let us first discuss the general properties of the solutions to the linear problem for the three-point function. The worldsheet describing the three-point function is a thrice-punctured sphere (see also figure \ref{fig3}) and one can define, around each puncture, a set of solutions $i_{\pm}$  $(i=1,2,3)$ with a definite monodromy property:
\begin{align}\label{eq:pipipi}
i_{\pm} \to e^{\pm ip_i (x)}i_{\pm} \,,\qquad p_i(x)\equiv \frac{2\pi \Delta_i x}{\sqrt{\lambda}(x^2-1)}\,.
\end{align}
To fix the normalization, we further impose the following conditions:
\begin{align}
&\langle i_{+}\,,i_{-}\rangle=1\,,\\
&g^{-1}\sigma_2i_{+}(x) =i \times i_{-}(1/x) \,,\qquad g^{-1}\sigma_2i_{-}(x) =i \times i_{+}(1/x)\,.
\end{align}

We expect that the classical string configuration approaches the simple solutions described in \eqref{simplesol2} near the positions of the vertex operators. This implies that the solutions to the linear problem can be approximated by the solutions $\psi_{\pm}$ with $\kappa$ and $\Delta$ replaced appropriately. Namely, we expect the following asymptotic behaviors of the solutions $i_{\pm}$: 
\begin{align}\label{eq:asymptoticformi}
\begin{aligned}
&i_{+}\overset{z\to z_i}{\sim} A_i(z-z_i)^{\frac{-\kappa_i}{1-x}}(\bar{z}-\bar{z}_i)^{\frac{-\kappa_i}{1+x}}\left( \begin{array}{c}-{\sf x}_i\\1\end{array}\right)\,,\\
&i_{-}\overset{z\to z_i}{\sim}- \frac{A_i^{-1}}{1+{\sf x}_i^2}(z-z_i)^{\frac{\kappa_i}{1-x}}(\bar{z}-\bar{z}_i)^{\frac{\kappa_i}{1+x}}\left( \begin{array}{c}1\\{\sf x}_i\end{array}\right)\,.
 \end{aligned}
\end{align}
Here $A_i$ is some constant which is to be determined later\footnote{One might think that such constants can be set to $1$ by rescaling the solutions $i_{\pm}\to (A_{i})^{\mp 1}i_{\pm}$. However, the rescaling in general is incompatible with our normalization condition $g^{-1}\sigma_2i_{\pm}(x) =i i_{\mp}(1/x)$. Therefore, in general, there is some unremovable overall factor $A_i$.}. 

Any solution to the linear problem can be expressed by a linear combination of these solutions as
\begin{align}
\psi=\langle \psi,i_{-}\rangle i_{+}-\langle \psi,i_{+}\rangle i_{-}\,.
\end{align}
Using this expansion and the reconstruction formula \eqref{reconAdS}, one can read off the asymptotic behavior of the string configuration $g^{-1}$ around each vertex operator:
\begin{align}
\begin{aligned}\label{behaviornearvertex}
g^{-1}&=\left. (\psi_1,\psi_2)\right|_{x=0}\overset{z\to z_i}{\sim}\left.(\langle \psi_1,i_{-}\rangle,\langle \psi_2,i_{-}\rangle)i_{+}\right|_{x=0}\\
&\sim \left.A_i|z-z_i|^{-2\kappa_i}\left(\begin{array}{cc}-\langle \psi_1,i_{-}\rangle{\sf x}_i&-\langle \psi_2,i_{-}\rangle{\sf x}_i\\\langle \psi_1,i_{-}\rangle&\langle \psi_2,i_{-}\rangle\end{array}\right)\right|_{x=0}\,.
\end{aligned}
\end{align}
(Here we used $i_{+}\gg i_{-}$ around $z=z_i$.) Now, comparing \eqref{behaviornearvertex} with the near boundary behavior of $g^{-1}$,\footnote{${\sf z}$ and ${\sf x}$ are the Poincare coordinates defined in \eqref{defPoinCare}.}
\begin{align}
g^{-1}=\left(\begin{array}{cc}X^{-1}-X^{4}&-X^{1}\\-X^{1}&X^{-1}+X^{4}\end{array}\right)\quad \overset{\text{near boundary}}{\to} \quad \frac{1}{{\sf z}}\left(\begin{array}{cc}{\sf x}_i^{2}&-{\sf x}_i\\-{\sf x}_i&1\end{array}\right)\,,
\end{align}
we can express the insertion points of the operators in terms of Wronskians:
\begin{align}\label{eq:differencecoor}
{\sf x}_i=\left.-\frac{\langle \psi_1 \,,i_{-}\rangle}{\langle \psi_2 \,,i_{-}\rangle}\right|_{x=0}\,.
\end{align}
One can then evaluate the right hand side of \eqref{eq:contributionfromvertex} as
\begin{align}
\begin{aligned}
&-\log {\rm Tr}\left[\left(\begin{array}{cc}1&{\sf x}_i\\{\sf x}_i&{\sf x}_i^{2}\end{array} \right)g^{-1}\right] (z_i) =\\
&-\log A_i-\log \left.\langle \psi_2,i_{-}\rangle\right|_{x=0} -\log \left(|z-z_i|^{-2\kappa_i}{\rm Tr}\left[\left(\begin{array}{cc}1&{\sf x}_i\\{\sf x}_i^2&{\sf x}_i\end{array}\right)\left(\begin{array}{cc}{\sf x}_i^{2}&-{\sf x}_i\\-{\sf x}_i&1\end{array}\right)\right]\right)\,,
\end{aligned}
\end{align}
Here we split the contribution into two terms; the first term $\langle \psi_2 , i_{-}\rangle$ is solution-dependent while the second term is universal and does not depend on a particular solution we consider. Furthermore, the second term is divergent since the trace inside the logarithm vanishes. As in the analysis of the Liouville field theory\footnote{See the discussion below \eqref{eq:discussiondiv}.}, such a divergence can be cancelled by renormalizing the vertex operator. We therefore obtain the following formula after the renormalization:
\begin{align}\label{eq:VfromW1}
\begin{aligned}
&\frac{\partial \log \langle\mathcal{O}_{\Delta_1}\mathcal{O}_{\Delta_2}\mathcal{O}_{\Delta_3} \rangle }{\partial \Delta_i}=-\log {\rm Tr}\left[\left(\begin{array}{cc}1&{\sf x}_i\\{\sf x}_i&{\sf x}_i^{2}\end{array} \right)g^{-1}\right] (z_i) =-\log A_i-\left.\log \langle \psi_2,i_{-}\rangle\right|_{x=0} \,.
\end{aligned}
\end{align}
 
 Although the formula \eqref{eq:VfromW1} allows us to express the contribution from the vertex operator in terms of the Wronskian, it still involves the solution $\psi_2$ which is unknown. To eliminate $\psi_2$, we first compute the difference of the coordinates using the formula \eqref{eq:differencecoor}:
 \begin{align}
 \begin{aligned}
 {\sf x}_i-{\sf x}_j=\left.-\frac{\langle \psi_1\,, i_{-}\rangle\langle \psi_2\,, j_{-}\rangle-\langle \psi_2\,, i_{-}\rangle\langle \psi_1\,, j_{-}\rangle}{\langle \psi_2\,, i_{-}\rangle\langle \psi_2\,, j_{-}\rangle}\right|_{x=0}\,.
 \end{aligned}
 \end{align}
 Using the identity for the epsilon tensor,
 \begin{align}
 \epsilon^{ab}\epsilon^{cd}+\epsilon^{ad}\epsilon^{bc}+\epsilon^{ac}\epsilon^{db}=0\,,
 \end{align}
 one can show\footnote{This identity is called the Schouten identity or the Plucker relation.}
 \begin{align}
 \langle \psi_1\,, i_{-}\rangle\langle \psi_2\,, j_{-}\rangle-\langle \psi_2\,, i_{-}\rangle\langle \psi_1\,, j_{-}\rangle=\langle \psi_1\,,\psi_2\rangle \langle i_{-}\,,j_{-}\rangle=\langle i_{-}\,,j_{-}\rangle\,.
 \end{align}
 We thus arrive at the expression
 \begin{align}
 {\sf x}_i-{\sf x}_j=-\left.\frac{\langle i_{-}\,, j_{-}\rangle}{\langle \psi_2\,, i_{-}\rangle\langle \psi_2\,, j_{-}\rangle}\right|_{x=0}\,.
 \end{align}
 One can then express the Wronskian involving $\psi_2$ in terms of the Wronskians of $i_{-}$'s as
 \begin{align}
\langle \psi_2\,,i_{-}\rangle^2= \frac{({\sf x}_j-{\sf x}_k)}{({\sf x}_i-{\sf x}_j)({\sf x}_i-{\sf x}_k)}\frac{\langle i_{-}\,,j_{-}\rangle\langle k_{-}\,,i_{-}\rangle}{\langle j_{-}\,,k_{-}\rangle}\,.
 \end{align}
 Plugging in this formula to \eqref{eq:VfromW1}, we get
 \begin{align}\label{eq:VfromW2}
\begin{aligned}
&\frac{\partial \log \langle\mathcal{O}_{\Delta_1}\mathcal{O}_{\Delta_2}\mathcal{O}_{\Delta_3} \rangle }{\partial \Delta_i}=-\log A_i-\frac{1}{2}\log \left[\frac{({\sf x}_j-{\sf x}_k)}{({\sf x}_i-{\sf x}_j)({\sf x}_i-{\sf x}_k)}\left.\frac{\langle i_{-}\,,j_{-}\rangle\langle k_{-}\,,i_{-}\rangle}{\langle j_{-}\,,k_{-}\rangle}\right|_{x=0}\right] \,.
\end{aligned}
\end{align}
 As the formula shows, we still need to figure out what $A_i$ is. This can be achieved in the following way. First consider the expression \eqref{eq:asymptoticformi} and take $x\to \infty$. Then, the solution to the linear problem becomes independent of the worldsheet coordinates and we obtain
 \begin{align}
 i_{+}(x=\infty) =A_i\left( \begin{array}{c}-{\sf x}_i\\1\end{array}\right)\,.
 \end{align}
 It is then straightforward to show that the following equality holds:
 \begin{align}
 A_i^{2} &= \frac{({\sf x}_j-{\sf x}_k)}{({\sf x}_i-{\sf x}_j)({\sf x}_k-{\sf x}_i)}\left.\frac{\langle i_{+},j_{+} \rangle\langle k_{+},i_{+} \rangle}{\langle j_{+},k_{+} \rangle}\right|_{x=\infty}\\
 &= \frac{({\sf x}_j-{\sf x}_k)}{({\sf x}_i-{\sf x}_j)({\sf x}_k-{\sf x}_i)}\left.\frac{\langle i_{-},j_{-} \rangle\langle k_{-},i_{-} \rangle}{\langle j_{-},k_{-} \rangle}\right|_{x=0}\,.
 \end{align}
 Here on the second line, we used the $\mathbb{Z}_2$ symmetry \eqref{eq:Z2symmetrypsi} and also the invariance of the Wronskian under the SL(2) transformations.
 
 As a result, we finally get the following expression for the contribution from the vertex operator:
  \begin{align}\label{eq:VfromW3}
\begin{aligned}
&\frac{\partial \log \langle\mathcal{O}_{\Delta_1}\mathcal{O}_{\Delta_2}\mathcal{O}_{\Delta_3} \rangle }{\partial \Delta_i}=-\log \left[\frac{({\sf x}_j-{\sf x}_k)}{({\sf x}_i-{\sf x}_j)({\sf x}_i-{\sf x}_k)}\left.\frac{\langle i_{-}\,,j_{-}\rangle\langle k_{-}\,,i_{-}\rangle}{\langle j_{-}\,,k_{-}\rangle}\right|_{x=0}\right] \,.
\end{aligned}
\end{align}
Therefore, the computation of the three-point function boiled down to the computation of the Wronskians $\langle i_{-},j_{-} \rangle$.
\subsection{Functional equation for the Wronskian}
To compute the Wronskians, we follow the same steps as the ones we did for the Liouville theory. Namely, we first derive the functional equation and solve it. The derivation of the functional equation is identical to the one in section \ref{subsec:wronsfunc}, and we only display the final result:
\begin{align}\label{eq:tosolvewron}
\langle i_{-},j_{-}\rangle(x)\langle i_{-},j_{-}\rangle(1/x)=\frac{\sin \left(\frac{p_i+p_j+p_k}{2}\right)\sin \left(\frac{p_i+p_j-p_k}{2}\right)}{\sin p_i \sin p_j}\,.
\end{align}
Here $p_i$'s are the quasi-momenta given in \eqref{eq:pipipi}. To solve this functional equation, let us first consider a simpler equation
\begin{align}\label{eq:toyfunctional}
k(x)+k(1/x) =h(x)\,,
\end{align}
where we assume that $h(x)$ is invariant under $x\to 1/x$.
A formal solution to this functional equation is given by
\begin{align}
\left. k(x)\right|_{|x|<1}=\oint_{|x^{\prime}|=1}\frac{dx^{\prime}}{4\pi i}\frac{k(x^{\prime})(1-x^2)}{(x-x^{\prime})(1-x x^{\prime})}\,.
\end{align}
Here the solution is defined for the unit disk $|x|<1$. To read off the value outside this unit disk, one has to perform the anlalytic continuation of the expression on the right hand side. Upon doing so, the poles\footnote{Note that there are two poles $x^{\prime}=x$ and $x^{\prime}=1/x$. The former pole crosses the contour from the inside to the outside while the latter pole crosses the contour from the outside to the inside.} of the integrand cross the contour and we get the following expression
\begin{align}
\begin{aligned}
\left. k(x)\right|_{|x|>1}=&\frac{h(x)+h(1/x)}{2}+\oint_{|x^{\prime}|=1}\frac{dx^{\prime}}{4\pi i}\frac{k(x^{\prime})(1-x^2)}{(x-x^{\prime})(1-x x^{\prime})}\,\\
=&h(x)+\oint_{|x^{\prime}|=1}\frac{dx^{\prime}}{4\pi i}\frac{k(x^{\prime})(1-x^2)}{(x-x^{\prime})(1-x x^{\prime})}\,.
\end{aligned}
\end{align}
  It is now straightforward to check that the solution satisfies \eqref{eq:toyfunctional}. For instance, assuming $|x|>1$, we get
  \begin{align}
  \begin{aligned}
  &k(x)+k(1/x)=\\
  &h(x)+\oint_{|x^{\prime}|=1}\frac{dx^{\prime}}{4\pi i}\frac{k(x^{\prime})(1-x^2)}{(x-x^{\prime})(1-x x^{\prime})}+\oint_{|x^{\prime}|=1}\frac{dx^{\prime}}{2\pi i}\frac{k(x^{\prime})(1-x^{-2})}{(x^{-1}-x^{\prime})(1-x^{-1} x^{\prime})}=h(x)\,.
  \end{aligned}
  \end{align}
  Note that, as we mentioned already in the first lecture, the solution to the functional equation is not unique owing to the ambiguities coming from the CDD factors. In the first lecture, we fixed such ambiguities by comparing the solution with the WKB expansion. Unfortunately, the WKB expansion turned out to be less powerful here and will not fix the ambiguity completely. Instead, we need to carefully discuss the analytic properties of the Wronskians in order to determine the CDD ambiguity\footnote{Note that in general the multiplication of the CDD-like factors will modify the analytic property of the Wronskians (as a function of the spectral parameter). Therefore, if one can determine the analytic property of the Wronskian by some other means, one will be able to fix the CDD ambiguities.}. This was for instance demonstrated in \cite{KKN}. However, since the discussion is quite technical, we will not discuss the CDD ambiguities in this lecture.
  
Now, applying the same technique to the logarithm of \eqref{eq:tosolvewron}, we get the following formal solution to the functional equation,
\begin{align}
\begin{aligned}
&\left. \log \langle i_{-},j_{-}\rangle\right|_{|x|<1}= \\
&\qquad {\tt CDD}+f\left[\frac{p_1+p_2+p_3}{2}\right]+f\left[\frac{p_i+p_j-p_k}{2}\right]-f[p_i]-f[p_j]\,,
\end{aligned}
\end{align}
with
\begin{align}\label{logsin}
f[h](x)\equiv \oint_{|x^{\prime}|=1}\frac{dx^{\prime}}{4\pi i}\frac{\log \sin h(x^{\prime})(1-x^2)}{(x-x^{\prime})(1-x x^{\prime})}\,.
\end{align}
Here ${\tt CDD}$ denotes the CDD ambiguities which we did not fix in the lecture\footnote{A more detailed analysis in \cite{KKN} actually shows ${\tt CDD}=0$.}.

Having determined the Wronskian, now it is straightforward to use the equality to compute the three-point function. The result is then given by
\begin{align}
\langle \mathcal{O}_{\Delta_1}\mathcal{O}_{\Delta_2}\mathcal{O}_{\Delta_3}\rangle=\frac{C_{123}}{{\sf x}_{12}^{\Delta_1+\Delta_2-\Delta_3}{\sf x}_{23}^{\Delta_2+\Delta_3-\Delta_1}{\sf x}_{31}^{\Delta_3+\Delta_1-\Delta_2}}
\end{align}
where $C_{123}$ is given by
\begin{align}\label{finalAdS}
\log C_{123}=\sum_{i}F[2 p_i]-\left(F[p_1+p_2+p_3]+\sum_{i\neq j\neq k}F[p_i+p_j-p_k]\right)+{\tt CDD}+{\tt S^5}
\end{align}
\begin{align}\label{mudeformed}
F[h(x)]\equiv \oint_{|x|=1}du\,\,{\rm Li}_2 \left( e^{i h(x)}\right)\,.
\end{align}
with
\begin{align}
u(x)=\frac{\sqrt{\lambda}}{4\pi}\left( x+\frac{1}{x}\right)\,.
\end{align}
This is the final result for the classical three-point function in AdS$_2$. As indicated, to compute the full three-point function, one also needs to take into account the contribution coming from $S^5$ part. This is significantly more complicated since the motion on the $S^5$ part knows the details of the operators and can be quite complicated in general. See \cite{KKN} for details and the final result. 

Although not complete, the result \eqref{finalAdS} already exhibits important features of the full answer. It is given in terms of certain combination of functions and each function is given by integration of the dilogarithm on the spectral curve. Interestingly, a similar dilogarithm was also obtained in the semi-classical limit at weak coupling \cite{Tailoring,Ivan}. This was one of the important evidence for the existence of the integrable structure threading both weak and strong coupling regimes\footnote{Recently, part of the classical string result was rederived from the non-perturbative approach called the hexagon formalism in \cite{clustering}.}. 

\subsubsection*{Some curious observations}
Before closing this section, let us mention some curious observations regarding the function \eqref{logsin}. Although not obvious, it turns out that the function \eqref{logsin} is essentially equivalent to the so-called {\it $\mu$-deformed Gamma function}, introduced in \cite{ppwave} as the ``massive" generalization of the Gamma function\footnote{It was recently shown that functions with similar properties can be obtained from the hexagon formalism at strong coupling \cite{Basso}}. Furthermore, the same function appears also in the {\it twistorial} generalization\footnote{In \cite{CNV}, they also introduced the twistorial generalization of the DOZZ formula. This is closely related to our final formula \eqref{finalAdS}.} of the topological string in \cite{CNV}. All these are suggesting that the formula \eqref{finalAdS} is a natural generalization of the Liouville three-point function, which is expressed in terms of ordinary Gamma functions. It would be interesting to understand deeper meaning of this coincidence.
\section{Conclusions and speculations}
In these lectures, we have seen that the idea akin to the ODE/IM correspondence can be effectively applied to study the three-point functions in both Liouville theory and string theory in AdS$_2$. Besides the similarity in methods, the structures of the final results (\eqref{finalliouville} and \eqref{finalAdS}) also resemble with each other: Both of them are expressed in terms of simple function(al)s and the combinations which appear in the arguments are essentially the same. It would be important to understand the origin of this similarity in more detail.

Asking this question is in particular of interest in view of the fact that the quantum three-point function in Liouville theory also exhibits the same structure. One derivation of the quantum three-point function in Liouville theory is due to Teschner \cite{Teschner,Teschner2} and it utilizes the crossing kernel (which is often called the fusion matrix \cite{MooreSeiberg}) which relates the conformal blocks in one channel to another. It turns out that this derivation shares many features with the computation of classical three-point functions presented in this lecture. For instance, the Schr\"{o}dinger equation is nothing but the classical limit of the BPZ differential equation\footnote{It is worth noting that a counterpart of the BPZ differential equation on the worldsheet theory was identified in \cite{Dei} for string theory on $AdS_3$ with NSNS fluxes.} for a degenerate operator, and the Wronskians $\langle i_{\pm},j_{\pm}\rangle$ can be regarded as the classical counterparts of the crossing kernel. Recently in \cite{openVerlinde}, the crossing kernels involving degenerate operators were re-interpreted as an open-path version of the Verlinde line operators. A natural next question is whether one can generalize it to string theory on $AdS_5\times S^5$. This would provide a new framework to study correlation functions in $\mathcal{N}=4$ SYM \cite{SK}. 

There are also things to be done on the Liouville side of our story. First of all, it would be interesting to generalize our analysis to the four-point function and derive a functional relation governing the classical conformal blocks. For strings in AdS$_2$, this was already achieved in \cite{CT}, and it would not be so difficult to perform the same analysis in the Liouville theory since it is actually simpler\footnote{See also a recent beautiful work on a related subject \cite{CGL} and references therein.}. It is also interesting to study correlation functions of irregular vertex operators \cite{GT}. This would allow us to make a more direct connection with the standard ODE/IM correspondence.  (See also \cite{Marino, opers} for related analysis.) Another possibility is to study disk one-point functions \cite{boundaryL1,boundaryL2}\footnote{Also interesting would be to analyze the disk one-point functions in classical string theory in $AdS_5\times S^{5}$. This will be relevant for the recent works on one-point functions in the presence of domain wall defects \cite{dL1,dL2,dL3,dL4,dL5}, and the three-point functions of two determinant operators and a single-trace operator \cite{JKV1,JKV2}.}. 

\end{document}